\documentclass{new_tlp}
\usepackage{mathptmx}
\usepackage{amsmath}
\usepackage{xspace}
\usepackage{color}
\usepackage{tikz}
\usetikzlibrary{positioning, automata}
\usepackage{verbatim}
\usepackage{natbib}
\tikzset{place/.style={circle, draw}}
\tikzset{>=stealth, auto, node distance=2.5cm, every loop/.style={->, min distance=10mm, in=0, out=60, looseness=10}}

\def\smpf{\hbox{SM$_{\bf p}[F]$} }

\def\bp{\textbf{p}}

\def\mp{{P}}
\def\mh{\hbox{${P_{hc}}$}}
\def\ar{\leftarrow}
\def\rar{\rightarrow}
\def\lrar{\leftrightarrow}
\def\beq{\begin{equation}}
\def\eeq#1{\label{#1}\end{equation}}
\def\ba{\begin{array}}
\def\ea{\end{array}}
\def\ii#1{\hbox{\it #1\/}}

\def\no{\naf}

\def\hc{{\em Hamiltonian Cycle}\xspace}

\def\definition{def-module\xspace}
\def\definitions{{\definition}s\xspace}
\newcommand{\vect}[1]{\bf{{#1}}}
\newcommand{\citeyearp}[1]{(\citeyear{{#1}})}

\def\naf{\hbox{\em not }}

\def\choice#1{\ii{Choice}(#1)}

\def\false{\hbox{\bf f}}
\def\true{\hbox{\bf t}}

\def\bx{\hbox{\bf x}}

\def\pred{\pi}
\def\formula{\mathcal{F}}

\def\shift{\hbox{$\mathit{shift}$}}

\newtheorem{prop}{Proposition}

\newtheorem{theorem}{Theorem}
\newtheorem{observation}{Observation}

\newtheorem{cor}{Corollary}

\title[First-Order Modular Logic Programs and their Conservative Extensions]{First-Order 
Modular Logic Programs and\\ their Conservative Extensions}
\author[Harrison and Lierler]{AMELIA HARRISON\\ 
University of Texas at Austin\\
\email{ameliaj@cs.utexas.edu}
\and YULIYA LIERLER\\
University of Nebraska  Omaha\\
\email{ylierler@unomaha.edu}
}

\begin{document}

\date{}
\maketitle

\begin{abstract}
Modular logic programs provide a way of viewing logic programs 
as consisting of many independent, meaningful modules. This paper 
introduces first-order modular logic programs, which can capture the 
meaning of many answer set programs. We also introduce conservative
extensions of such programs. This concept helps to identify strong 
relationships between modular programs as well as between traditional 
programs. We show how the notion of a conservative extension can be
used to justify the common projection rewriting. This note is under 
consideration for publication in Theory and Practice of Logic Programming. 
\end{abstract}

\section{Introduction}
Answer set programming (ASP) is a prominent knowledge representation paradigm rooted 
in logic programming. In ASP, a software developer represents a given computational problem by a
program whose answer sets (also called stable models) correspond to solutions. Then, the developer uses an 
answer set solver
to generate stable models for the program. 
%Answer sets serve as the main mean to grasp the meaning of a  program at hand. 
%Programs with schematic variables are ubiquitous in the practice of answer set 
%programming.  
In this paper we show how some logic programs can be viewed as consisting of 
various ``modules'', and how stable models of these programs can be computed by composing the 
stable models  of the modules. We call collections of such modules first-order modular programs. 
To illustrate this approach consider the following two rules
\begin{align}
&{\it r}(X, Y) \ar {\it in}(X,Y). \label{eq:rdef1}\\
&{\it r}(X,Y) \ar {\it r}(X,Z), {\it r}(Z,Y). \label{eq:rdef2}
\end{align}
%TODO: find a place for this!
%where we consider $Y\neq Z$ and $X\neq Y$ to be abbreviations for expressions  
%$not\ (Y= Z)$ and $not\ (X= Y)$, respectively.
Intuitively, these rules encode %the information 
that the relation ${\it r}$ is the transitive closure of the 
relation ${\it in}$. The empty set is the only answer set of the program 
composed of these rules alone. Thus, in some sense the meaning of these two 
rules in isolation is the same as the meaning of any program that has a single answer set that is empty. 
We show how we can view these rules as forming a 
module and use the operator SM introduced by Ferraris et al.~\citeyearp{fer09}  to define a 
semantics that corresponds more accurately to the intuition associated with the 
rules above. The operator SM provides a definition of the stable model semantics 
for first-order logic programs that does not refer to grounding or fixpoints as does the original definition. 
The operator SM has proved to be an effective tool for studying the properties 
of logic programs with variables. Since such programs are the focus of this paper, we  
 chose the operator SM as a technical tool here.

 Modularity is essential for modeling large-scale practical applications. Yet research on modular answer set programming is at an
 early stage.
 Here we propose first-order modular programs and argue their  
utility for reasoning about answer set programs. 
We use the \hc problem as a running example to illustrate that a ``modular" view 
of a program gives us  
\begin{itemize} 
	\item a more intuitive reading of the parts of the program;
	\item the ability to incrementally develop modules or parts of a program 
that have stand-alone meaning and that interface with other modules via a common signature; 
	\item a theory for reasoning about modular rewritings of individual 
components with a clear picture of the overall impact of such changes.\end{itemize}	
First-order modular programs introduced here can be 
viewed as a generalization of propositional 
 modular logic programs~\citep{lie13}. % to the first-order case. 
% The idea for the semantics of  modular programs is borrowed from 
% the definition of the semantics for propositional 
% modular logic programs, but in this work we cover a broader 
% (first-order) class of programs. 
In turn, propositional 
 modular logic programs generalize the concept of lp-modules by Oikarinen and Janhunen~\citeyearp{oik08}.
 ASP-FO logic~\citep{dltj12} is another related formalism. It is a modular 
formalization of {\em generate-define-test} answer set programming~\citep{lif02} 
that allows for unrestricted interpretations as
 models, non-Herbrand functions, and first-order formulas in the bodies of rules.
 An ASP-FO theory is a set consisting of modules of three types: G-modules 
 (G for generate), D-modules (D for define), and T-modules (T for test). In 
 contrast, there is no notion of type among modules in the  modular programs 
 introduced here.
 
We also define conservative extensions for first-order modular programs. 
This concept is related to strong equivalence for logic programs~\citep{lif01}.
If two rules are strongly equivalent, we can replace one with the other
within any program 
and the answer sets of the resulting program will coincide with those of 
the original one.
%in the context of first-order modular programs. 
Conservative extensions allow  
us to reason about rewritings even when the rules in question have 
different signatures.
We can justify the common projection rewriting described in \cite{fab99} using this concept. 
 For example, the rule 
\beq
\ar not\ {\it r}(X, Y), {\it edge}(X, Z), {\it edge}(Z', Y) 
\eeq{eq:rcon}
says that every vertex must be reachable from every other vertex.
This rule can be
replaced with the following three rules without affecting the stable models in 
an ``essential way'' 
$$
\ba{l}
\ar \no {\it r}(X, Y)\wedge {\it vertex1}(X) \land {\it vertex2}(Y).  \\
 {\it vertex1}(X) \ar {\it edge}(X,Y).\\
 {\it vertex2}(Y) \ar {\it edge}(X,Y).\\
\ea 
$$
Furthermore, this replacement is valid in the context of any program, as long as 
that program does not already contain either of the predicates ${\it vertex1}$ 
and ${\it vertex2}$. 
Such rewritings can be justified using conservative extensions. 
Conservative extensions provide a theoretical justification for rewriting 
techniques already commonly in use. Projection is one such  
technique, which often improves the performance of answer set 
programs. Currently, these performance-enhancing rewritings are done 
manually.
We expect the theory about conservative extensions developed here will provide 
a platform for automating such rewritings in the future. 	 	
We note that conservative extensions are related to the notion of knowledge 
forgetting in \cite{wang14}. However, that work applies only to propositional 
programs. 

This paper is structured as follows. In Sections \ref{sec:trpr} and 
\ref{sec:sm} we review traditional programs 
and the operator SM. In Section \ref{sec:modlog}, we define first-order modular 
logic programs, and in Section~\ref{sec:relprog} we show how they are related to 
traditional logic programs. Finally, in Section \ref{sec:eqre}, we introduce 
conservative extensions and show how they can be used to justify program rewritings.  

%Section~\ref{sec:sm} reviews the SM operator introduced by Ferraris et 
%al.~\citeyearp{fer09}, which provides an alternative to the definition of the answer set (stable model) semantics of logic programs considered in this section.
%The SM operator allows for non-empty sets of stable models for the program composed  
%of rules~\eqref{eq:rdef1} and~\eqref{eq:rdef2}, when predicate {\it r} is 
%considered as ``intensional'' and predicate {\it in} as ``extensional''. Such a 
%treatment of these rules gives us a more accurate correspondence between their intuitive reading and their model-theoretic semantics.

%Later, we will see that the \hc program can be viewed as a 
%``modular program'' and that doing so will allow us to swap parts of the program 
%with different sets of rules without changing its answer sets. 

\section{Review: Traditional Programs}\label{sec:trpr}
A {\em (traditional logic) program} is a finite set of {\em rules}  of the form
\beq
a_1 ; \dots ; a_k \ar a_{k+1} , \dots , a_l , not\  a_{l+1} , \dots 
, not\  a_m , not\  not\  a_{m+1} , \dots , not\  not\  a_n, 
\eeq{eq:ruletraditional}
$(0\leq k \leq l \leq m \leq n)$, where each $a_i$ is an atomic formula, 
possibly containing function symbols, variables, or the equality symbol with the restriction that 
atomic formulas $a_1, \dots, a_k$ and $a_{m+1}, \dots, a_{n}$  
may not contain the equality symbol.
The expression containing atomic 
formulas $a_{k+1}$ through $a_{n}$ is called the {\em body} of the rule. A rule 
with an empty body is called a {\em fact}.  
An {\em instance} of a rule~$R$ occurring in a program $\Pi$ is a rule that can be formed 
by replacing all variables occurring in~$R$ with ground terms formed from  
function symbols and object constants occurring in $\Pi$.  
The process of {\em grounding} a  traditional logic program consists of the 
following steps: 
\begin{enumerate}\setlength\itemsep{0em}
 \item each rule is replaced with 
all of its instances by substituting ground terms for variables;  
\item in each instance, every atomic formula of the form $t_1 = t_2$ is replaced 
by $\top$ if $t_1$ is the same as $t_2$ and by $\bot$ otherwise. 
%\item among all ground instances formed in the previous step, 
%\begin{enumerate}\setlength\itemsep{0em}
%\item  each occurrence of an atomic formulas $a_i$ is eliminated 
%if $k+1 \leq i \leq l$ and $a_i$ is of the form $t=t$, 
%\item  each occurrence of an expression $not\  a_i$ 
%is eliminated if $l+1 \leq i \leq m$  and $a_i$ is of the form $t=t'$  
%where $t$ and $t'$ are syntactically different objects; 
%\end{enumerate}
%\item after the above steps are complete all ground instances that contain equality symbols are eliminated.
\end{enumerate}
It is easy to see that the resulting ground program does not have equality 
symbols and can be viewed as a propositional program.
The {\em answer sets} of
a traditional program $\Pi$ are stable models of the result of grounding  
$\Pi$, where stable models are understood as in \citep{fer05}. 

According 
to~\citep{fer05b} and \citep{fer05}, rules of the form~\eqref{eq:ruletraditional} are sufficient 
to capture the meaning of the choice rule construct commonly used in 
answer set programming. For instance, the choice rule ${\tt \{p(X)\}\ar q(X)}$ 
is understood as the rule 
$$p(X)\ar q(X),\ not\ not\ p(X).$$
In this paper we adopt choice rule notation. 
Traditional logic programs cover a substantial practical fragment of the input 
languages used in developing answer set programming applications. %For example, 
%the \hc program from the previous section is a traditional program. 

Consider the traditional program consisting of the rule
\beq
s(X,Z) \ar p(Z), q(X,Y), r(X,Y) 
\eeq{eq:s}
and the facts 
\beq
p(2).~ q( 1,1).~ q(1,2).~q(2,2).~r(1,1).~ r(1,2).~ r(2,1).~ 
\eeq{eq:facts1}
Grounding this program results in eight instances of~\eqref{eq:s} and  
the facts in~\eqref{eq:facts1}.
%and the following rules
%$$
%\ba{l}
%s(1,1) \ar p(1), q(1,1), r(1,1).\\ 
%s(1,2) \ar p(2), q(1,1), r(1,1).\\ 
%s(1) \ar p(1), q(1,2), r(1,2).\\
%s(1) \ar p(1), q(1,2), r(1,2).\\ 
%s(1) \ar p(1), q(1,2), r(1,2).\\
%s(2) \ar p(2), q(2,1), r(2,1).\\ 
%s(2) \ar p(2), q(2,2), r(2,2).\\
%\ea
%$$
The only answer set of this program is 
\beq
\{p(2),  q(1,1), q(1,2), q(2,2),r(1,1), r(1,2), r(2,1), s(1,2)\}.
\eeq{eq:as1}

Consider the \hc problem on an undirected graph. 
This problem is often used to introduce answer set 
programming. 
 A \hc is a subset of the 
set of edges in a graph that forms a cycle going though each vertex exactly once.
A sample program that encodes this 
can be constructed by adding rules \eqref{eq:rdef1}, \eqref{eq:rdef2}, and \eqref{eq:rcon} to the 
following:
\begin{align}
&{\it edge}(a, a'). \;\; \dots \;\; {\it edge}(c, c'). \label{eq:hcfacts}\\
&{\it edge}(X, Y) \ar {\it edge}(Y,X). \label{eq:edgesym}\\
& \{{\it in}(X,Y)\} \ar {\it edge}(X,Y). \label{eq:indef}\\
&\ar {\it in}(X,Y), {\it in}(X,Z), Y\neq Z. \label{eq:incon1}\\
&\ar {\it in}(X,Z), {\it in}(Y,Z), X\neq Y. \label{eq:incon2}\\
&\ar {\it in}(X,Y), {\it in}(Y,X).\label{eq:incon3}
\end{align}
%TODO: find a place for this!
%where we consider $Y\neq Z$ and $X\neq Y$ to be abbreviations for expressions  
%$not\ (Y= Z)$ and $not\ (X= Y)$, respectively.

Each answer set of the 
\hc program above corresponds to a Hamiltonian cycle 
%The answer sets of the 
%\hc program above  are in 1-1 correspondence with the Hamiltonian cycles 
of the given graph, specified by facts~\eqref{eq:hcfacts}, so that the predicate {\it in} encodes these cycles.
If an atom ${\it in}(a,b)$ appears in an answer set it says that the edge between $a$ and $b$ is part of the subset forming the Hamiltonian cycle. 
Intuitively, 
\begin{itemize}
\setlength\itemsep{0em}
\item the facts in~\eqref{eq:hcfacts} 
 define a graph instance by listing its edges, and rule~\eqref{eq:edgesym} 
ensures that this ${\it edge}$ relation is symmetric (since we are dealing with 
an undirected graph); the vertices of the graph 
are implicit---they are objects that occur in the edge relation;\footnote{This 
precludes graphs that include isolated vertices, but such vertices can be safely 
ignored when computing Hamiltonian cycles.}
\item rule~\eqref{eq:indef} says that any edge may belong to a Hamiltonian cycle;
\item rules~\eqref{eq:incon1} and~\eqref{eq:incon2}  impose the restriction that 
no two edges in a Hamiltonian cycle may start or end at the same vertex, and 
rule~\eqref{eq:incon3} requires that each edge appears at most once in a 
Hamiltonian cycle (recall that ${\it in}(a,b)$ and ${\it in}(b,a)$ both encode 
the information that the edge between $a$ and $b$ is included in  a Hamiltonian cycle);
\item rules~\eqref{eq:rdef1} and~\eqref{eq:rdef2} define a relation {\it r} 
(reachable) that is the transitive closure of relation {\it in};
\item rule~\eqref{eq:rcon} imposes the restriction that every vertex in 
a Hamiltonian cycle 
must be reachable from every other vertex.
\end{itemize}
Groups of rules of the \hc program have clear intuitive meanings as shown above. 
Yet, considering these groups separately will not produce ``meaningful'' 
logic programs under the answer set semantics as discussed in the introduction. 
%For instance,
%the program consisting solely of rule~\eqref{eq:indef}
%has a single answer set that is empty.
In this paper, we show how we can view each of these groups of rules as a 
separate module, and then use the SM operator introduced by Ferraris et 
al.~\citeyearp{fer09}, along with a judicious choice of ``intensional'' and 
``extensional'' predicates to achieve  
a more accurate correspondence between the intuitive reading of the groups of rules and their model-theoretic semantics.

\section{Review: Operator SM}\label{sec:sm}

The SM operator introduced by Ferraris et al.~\citeyearp{fer09} 
gives 
a definition for the semantics of logic programs with variables different than 
that 
described in the previous section. The SM operator bypasses grounding and  
provides a mechanism for viewing groups of rules in a program as separate units 
or ``modules''. Consider rule~\eqref {eq:s}. 
Intuitively, we attach a meaning to this rule:
it
expresses that relation $s$ holds for a pair of objects when property~$p$ holds of 
the second  
object and some object is in relation~$q$ and relation~$r$ with 
the first object.  
A program 
consisting only of this rule has a single answer set that is empty, which is inadequate to capture these intuitions. 
Ferraris et al.~\citeyearp{fer09} partition predicate symbols of a program into 
two groups: ``intensional'' and ``extensional''. If the predicate $s$ 
is  the only intensional predicate in rule~\eqref {eq:s}, then the SM 
operator captures the intuitive meaning of this rule seen as a program.
%The SM operator also provides a natural treatment of equality symbol in a program.

We now review the operator SM following~\citep{fer09}.
The symbols
$ \bot,\land, \lor, \rar, \forall,$ and $\exists$ 
are viewed as primitives. The formulas $\neg F$ %, $F \lrar G$,
 and $\top$ are 
abbreviations for $F\rar\bot$
%, $(F \rar G) \land (G \rar F),$  
and $\bot \rar 
\bot$, respectively. 
If $p$ and $q$ are predicate symbols of arity~$n$ then $p \leq q$ is an 
abbreviation for the formula 
$
\forall {\bf x}(p({\bf x}) \rar q({\bf x})), 
$
where ${\bf x}$ is a tuple of variables of length $n$. If ${\bf p}$ and ${\bf 
q}$ are tuples $p_1, \dots, p_n$ and $q_1, \dots, q_n$ of predicate symbols then 
${\bf p} \leq {\bf q}$ is an abbreviation for the conjunction 
$$
(p_1 \leq q_1) \land \dots \land (p_n \leq q_n),
$$ 
and 
${\bf p} < {\bf q}$ is an abbreviation for 
$({\bf p} \leq {\bf q}) \land \neg ({\bf q} \leq {\bf p}).$ We apply the same 
notation to tuples of predicate variables in second-order logic formulas.   
If ${\bf p}$ is a tuple of predicate symbols $p_1, \dots, p_n$ (not including equality), and $F$ 
is a first-order sentence 
then SM$_{\bf p}[F]$ (called the {\em stable model operator with intensional 
predicates ${\bf p}$}) denotes the second-order sentence 
$$ F \land \neg \exists {\bf u}({\bf u} < {\bf p}) \land F^*({\bf u}), $$
where ${\bf u}$ is a tuple of distinct predicate variables $u_1, \dots, u_n$, and $F^*({\bf u})$ is defined recursively:
\vspace{-.3cm}
\begin{itemize}
\setlength\itemsep{0em}
%\item $\bot^*$ is $\bot$;
%\item $(t_1 = t_2)^*$ is $(t_1 = t_2)$ for any terms $t_1, t_2$; 
\item $p_i({\bf t})^*$ is $u_i({\bf t})$ for any tuple ${\bf t}$ of terms;
\item $F^*$ is $F$ for any atomic formula $F$ that does not contain members of 
$\bp$;%\footnote{This includes equality statements and the formula $\bot$.} 
\item $(F \land G)^*$ is $F^* \land G^*$;
\item $(F \lor G)^*$ is $F^* \lor G^*$;
\item $(F \rar G)^*$ is $(F^* \rar G^*) \land (F \rar G) $;
\item $(\forall x F)^*$ is $\forall x F^*$;
\item $(\exists x F)^*$ is $\exists x F^*$.
\end{itemize}
\vspace{-.3cm}
Note that if ${\bf p}$ is the empty tuple then SM$_{\bf p}[F]$ is equivalent to 
$F$. For intuitions regarding the definition of the SM operator we direct the 
reader to \cite[Sections  2.3, 2.4]{fer09}. 

A {\em signature} is a set of function and 
predicate symbols.   A function symbol of arity 0 is an {\em object constant}. 
For an interpretation~$I$ over signature $\sigma$ and a function symbol (or, 
predicate symbol) $t$ from $\sigma$ by $t^I$ we denote a function (or, relation) 
assigned to $t$ by $I$.    Let  $\sigma$ and $\Sigma$ be  signatures so that 
$\sigma\subset\Sigma$.
For interpretation $I$ over  $\Sigma$, by $I_{|\sigma}$ we denote the 
interpretation over $\sigma$ constructed from $I$ so that for every function or 
predicate symbol $t$ in $\sigma$, $t^I=t^{I_{|\sigma}}$.

By $\sigma(F)$ we denote the the set of all 
function and 
predicate symbols occurring in formula $F$ (not including equality).
We will call this the {\em signature of $F$}. 
An interpretation $I$ over $\sigma(F)$
 is a ${\bf p}$-stable model of $F$ if it satisfies SM$_{\bf p}[F]$, where {\bf p} is a tuple of predicates from $\sigma(F)$. We 
will sometimes refer to ${\bf p}$-stable models where ${\bf p}$ denotes a {\emph set} rather than a tuple 
of predicates. Since the cardinality of ${\bf p}$ will always be finite, the 
meaning should be clear. It is easy to see that any ${\bf p}$-stable model of $F$ is also a model of $F$. 
Similarly, it is clear that for any interpretation $I$, if $I_{|\sigma(F)}$ is a 
${\bf p}$-stable model of $F$ then $I$ satisfies SM$_{\bf p}[F]$. We may 
refer to such an interpretation as a ${\bf p}$-stable model as well. 

From this point on, we view logic program rules as alternative notation for particular 
types of first-order 
sentences.  
For example,  rule~\eqref{eq:s}
is seen as an abbreviation for the first-order sentence 
\beq
\forall xyz((p(z) \land q(x,y) \land r(x,y)) \rar s(x,z)).
\eeq{eq:s1def}
Similarly, we understand the \hc program presented in Section~\ref{sec:trpr} as 
an abbreviation for the conjunction of the following formulas
\beq
 \ba{ll}
%\begin{align*}
&{\it edge}(a, a')\wedge \;\; \dots \;\; \wedge {\it edge}(c, c') \; \\
&\forall x y({\it edge}(y,x) \rar {\it edge}(x,y)) \; \\
&\forall x y( (\neg\neg  {\it in}(x,y) \wedge {\it edge}(x,y)) \rar {\it 
in}(x,y)) \; \\
&\forall x y z(({\it in}(x,y)\wedge {\it in}(x,z)\wedge \neg(y=z) )\rar\bot) 
\; \\
&\forall x y z(({\it in}(x,z)\wedge {\it in}(y,z)\wedge \neg(x=y)) \rar\bot) 
\; \\
&\forall x y(({\it in}(x,y)\wedge {\it in}(y,x))\rar\bot) 
\; \\
&\forall x y({\it in}(x,y) \rar {\it r}(x, y)) \; \\
&\forall x y z(({\it r}(x,z) \wedge {\it r}(z,y))\rar {\it r}(x,y)) \; \\
&\forall x y z z'((\neg {\it r}(x, y)\wedge {\it edge}(x, z)\wedge {\it 
edge}(z', y))\rar \bot) \\ 
\ea
\eeq{eq:hc2}
%\end{align*}
where $a,a',\dots c,c'$ are object constants and $x, y, z, z'$ are variables.\footnote{In 
logic programming it is customary to use uppercase letters to denote variables. 
In the literature on logic it is the specific letter used that indicates 
whether a symbol is  an object constant or a variable (with letters drawn from the 
beginning of the alphabet typically used for the former and letters from the end 
of the alphabet for the latter). We utilize both of these traditions 
depending on the context. %whether we use the syntax stemming from logic programming or logic.
}

Let $S$ denote sentence~\eqref{eq:s1def}.
We now illustrate the definition of ${\bf p}$-stable models.  If $s$ is the only 
intensional predicate occurring in $S$ then 
$S^*(s)$ is 
$$
\forall xyz(((p(z) \land q(x,y) \land r(x,y)) \rar u(x,z)) \land 
((p(z) \land q(x,y) \land r(x,y)) \rar s(x,z)))
$$ 
and SM$_{s}[S]$ is 
$$\ba{l}
S %\forall xyz((p(z) \land q(x,y) \land r(x,y)) \rar s(x,z)) 
\land %\\
\neg \exists u (( \forall xz (u(x,z) \rar s(x,z)) \land \neg \forall xz (s(x,z) \rar 
u(x,z))) \land S^*(s) % \\
%\forall xyz(((p(z) \land q(x,y) \land r(x,y)) \rar u(x,z)) \land 
%((p(z) \land q(x,y) \land r(x,y)) \rar s(x,z))).
\ea$$
This second-order sentence is equivalent to the first-order sentence
$$
\forall xz(s(x,z) \lrar (p(z) \land  \exists y (q(x,y) \land r(x,y)))),
$$
which reflects the intuitive meaning of the rule (\ref{eq:s}) seen as a program.

By $\pred(F)$ we denote the set of all 
predicate symbols (excluding equality) occurring in $F$.
The following theorem is slight generalization of Theorem 1 from \citep{fer09} 
as we consider quantifier-free formulas that may contain 
equality.

\begin{theorem}\label{thm:sm-traditional}
 Let $\Pi$ be a traditional logic program. If $\sigma(\Pi)$ contains at least one object constant then for any Herbrand interpretation $X$ of $\sigma(\Pi)$
 the following conditions are equivalent
 \begin{itemize}
\setlength\itemsep{0em}
\item  $X$ is an answer set of $\Pi$;
\item $X$ is a $\pred(\Pi)$-stable model of $\Pi$.   
\end{itemize}
\end{theorem}

This theorem illustrates that  the
set of Herbrand $edge,r,in$-stable models of program~\eqref{eq:hc2} 
coincide with the set of its answer sets.

\section{Modular Logic Programs}\label{sec:modlog}

In this section, we introduce 
first-order modular logic programs,
which are similar to the propositional 
modular logic programs introduced in~\citep{lie13}. In a nutshell, a first-order modular logic program is a collection of logic 
programs, where the SM operator is used to compute models of each individual logic program in the collection.
The semantics of a modular program is computed by finding the ``intersection'' of the 
interpretations that are models of its components. 
We call any formula of the form SM$_{\bf p}[F]$, where ${\bf p}$ is a tuple of 
predicate symbols and $F$ is traditional logic program viewed as a first-order 
formula,  a {\em defining module (of ${\bf p}$ in $F$)} or a {\em \definition}.  
A {\em first-order modular logic program} (or, {\em  modular program}) $\mp$ is a 
finite set of 
\definitions
%\beq
\[\{{\text SM}_{\bf p_1}[F_1], \dots , 
{\text SM}_{\bf p_n}[F_n]\}. 
\]%\eeq{eq:modprog} 
Let $\mp$ be a modular program. By 
$\sigma(\mp)$ we denote the set 
$$
\bigcup_{{\text SM}_{\bf p}[F]\in \mp} \sigma(F),
$$
called the {\em signature of $\mp$}. 
We say that an interpretation $I$ over the signature $\sigma(\mp)$ is a {\em 
stable model} of  modular program $\mp$ if 
for every \definition SM$_{\bf p}[F]$ in $\mp$,
$I_{|\sigma(F)}$ is a $\vect{p}$-stable model of $F$.

Let $P,Q,$ and~$R$ stand for formulas  
\beq
p(2),
\eeq{eq:p}
\beq
q(1,1) \land q(1,2) \land q(2,2), \text{ and}
\eeq{eq:q}
\beq
r(1,1) \land r(1,2) \land r(2,1),
\eeq{eq:r}
respectively. 
Consider a modular program consisting of four {\definition}s
\beq\{{\text SM}_p[P],
{\text SM}_q[Q],
{\text SM}_r[R], 
{\text SM}_s[S]\},
\eeq{eq:modprog1}
where $S$ is defined as in the previous section. 
The Herbrand interpretation~\eqref{eq:as1} 
is a stable model of this modular program. 

The stable models of  
modular program~\eqref{eq:modprog1} coincide with the $p,q,r,s$-stable models of 
\beq
\text{SM}_{p,q,r,s}[P\land Q \land R \land S].
\eeq{eq:sm_prog} 
Recall that $P \land Q \land R \land S$ can be viewed as the logic program consisting 
of the facts~\eqref{eq:facts1}
and the rule \eqref{eq:s}. By Theorem~\ref{thm:sm-traditional}, the Herbrand $p,q,r,s$-stable models 
of \eqref{eq:sm_prog} coincide with the answer sets of the logic program composed of rules in~\eqref{eq:s}
and~\eqref{eq:facts1}.
These facts hint at the close relationship between modular logic programs and 
traditional logic programs as written by answer set programming practitioners. 
In the following, we formalize the relationship between modular logic programs 
and traditional logic programs. This formalization is rooted in prior work on 
splitting logic programs from \cite{fer09a}.  

\section{Relating Modular Programs and Traditional Programs}\label{sec:relprog}

%To simplify presentation, we restrict our attention to 
%first-order sentences corresponding to traditional logic programs. 
As mentioned earlier, we view a traditional logic program
as an abbreviation for a first-order sentence formed as a conjunction of 
formulas of the form 
\beq
\widetilde{\forall} (a_{k+1} \land \dots \land a_l \land \neg a_{l+1} \land \dots 
\land \neg a_m \land \neg \neg a_{m+1} \land \dots \land \neg \neg a_n \rar a_1 
\lor \dots \lor a_k), 
\eeq{eq:rule}
which corresponds to rule~\eqref{eq:ruletraditional}. The symbol $\widetilde{\forall}$ denotes universal closure. 
We call the disjunction in the consequent of a rule~\eqref{eq:rule} its {\em 
head}, and the conjunction in the antecedent its {\em body}.
The 
conjunction $a_{k+1} \land \dots \land a_{l}$ constitutes the {\em positive part of the body}.
It is sometimes convenient to abbreviate the body of a rule with the letter $B$ 
and represent rule~\eqref{eq:rule} as 
\beq
\widetilde{\forall} (B \rar a_1 \lor \dots \lor a_k).
\eeq{eq:ruleabbr}

%We now introduce some notation used in the sequel.
Let $\mp$ denote a  modular program.
By 
$\pred(\mp)$ we denote the set 
$$
\bigcup_{{\text SM}_{\bf p}[F]\in \mp} \pred(F),
$$
called the {\em predicate signature} of $\mp$. 
Similarly, by $\iota(\mp)$ we denote the set 
$$
\bigcup_{{\text SM}_{\bf p}[F]\in \mp} {\bf p}
$$
called the {\em intensional signature} of  
$\mp$. By $\formula(\mp)$ we denote the formula 
$$
\bigwedge_{{\text SM}_{\bf p}[F]\in \mp}  F.
$$

%For example, 
%first-order formula $S$ from Section \ref{sec:sm} is a rule. 
%It is easy to see 
%that disjunctive logic rules as introduced in~\citep{gel91b}  can be represented using normal modular logic 
%programs. In fact, choice rules can also be represented in this syntax. For instance, we view 
%the choice rule ${\tt \{p(X)\}\ar q(X).}$ as an abbreviation for the rule $\forall x (q(X)\wedge \neg \neg p(x) \rar 
%p(x)).$

A modular program is called {\em simple} when % (i) it consists only of \definitions 
%${\text SM}_{\bf p}[F]$ where $F$ is a tradition%al logic program (viewed as a 
%first-order sentence) and (ii) 
 for every \definition ${\text SM}_{\bf p}[F]$,  
every predicate symbol $p$ occurring in the head of 
a rule in $F$ occurs also in the tuple $\bf p$.  For instance, modular 
program~\eqref{eq:modprog1} is simple.
We note that this restriction %(ii) 
%of a traditional modular programs 
is, in a sense, inessential. Indeed, consider  a \definition ${\text SM}_{\bf 
p}[F]$ that is not simple.
%$F$ satisfies restriction (i) but not (ii). 
There is a straightforward  syntactic 
transformation that can be performed on each rule in $F$, 
resulting in a formula $F'$ such that ${\text SM}_{\bf p}[F]$ is equivalent to ${\text SM}_{\bf p}[F']$. 
Let~$R$ be a rule of the form~\eqref{eq:ruleabbr} and {\bf p} be a tuple of predicate symbols. 
By $\shift_{\bf p}(R)$ we denote the universal closure of the following formula 
%\beq
\[
B  \;\; \land \bigwedge_{\pred(a_i)\not\in {\bf p}, \atop 1 \leq i \leq 
k } \neg a_i 
\;\;\rar \;\;\bigvee_{\pred(a_i)\in {\bf p}, \atop 1 \leq i \leq k } a_i.
\]
%\eeq{eq:ruleshift}
In other words,  any atomic formula in the head of a rule whose predicate symbol is not in~{\bf p} is moved to the body of 
the rule and preceded by negation. For a traditional logic 
program~$F$, $\shift_{\bf p}(F)$ is the conjunction of formulas obtained by 
applying~$\shift_{\bf p}$ to each rule in~$F$. Theorem~5 from \cite{fer09} shows that
if the equivalence between any two first-order formulas can 
be derived intuitionistically from the law of excluded middle formulas for all 
extensional predicates occurring in those formulas, then they have the same stable 
models. The following observation is a consequence of that theorem. 
\begin{observation}\label{thm:shift} 
For  a traditional logic program $F,$
\definitions  ${\text SM}_{\bf p}[F]$ and 
 ${\text SM}_{\bf p}[\shift_{\bf p}(F)]$
 are equivalent. 
\end{observation}

%For any rule~\eqref{eq:rule}, 
%We say that a predicate symbol $p$ is a {\em head} symbol of a 
%rule~\eqref{eq:rule} if it occurs in its head. Similarly,  a predicate symbol 
%$p$ is a {\em positive} symbol of a rule~\eqref{eq:rule} if it occurs in the 
%positive part of the body.
For any simple modular program $\mp$, 
the {\em dependency graph} of $\mp$, denoted $DG[\mp]$,
 is a directed graph that 
\begin{itemize}
\item has all members of the intensional signature $\iota(\mp)$  as its vertices, and
\item has an edge from $p$ to $q$ if there is a \definition 
${\text SM}_{\bf p}[F]\in \mp$ 
containing a rule with $p$ occurring in the head and~$q$ occurring in the positive part of the body.   
%\item has an edge from $p$ to $q$ if for some rule $r$ in $F$ so that ${\text SM}_{\bf p}[F]\in \mp$,
%$p$ is a head symbol in $r$, whereas $q$ is a positive symbol in $r$.   
\end{itemize}
For instance, the dependency graph of simple modular program~\eqref{eq:modprog1} 
consists of four vertices $p,q,r,s$ and edges from $s$ to $p$, from $s$ to~$q$, and from $s$ to $r$.
It is easy to see that this graph has four strongly connected components, each consisting of a single vertex.

%We say that a traditional formula $F$ is {\em negative} on a tuple {\bf p} of 
%predicate constants if no member of {\bf p} occurs in the head of any rule in 
%$F$.\footnote{A similar definition of predicate dependency graph
%was given in~\citep{fer09a} for sentences  of a more general syntax.} 

We call a simple modular program $\mp$
{\em coherent} if 
\begin{enumerate}
 \item[(i)] 
 for every pair of  distinct \definitions
 ${\text SM}_{\bf p}[F]$
 and 
 ${\text SM}_{\bf p'}[F']$ in $\mp$, tuples  ${\bf p}~\cap~{\bf p'}~=~\emptyset$, 
and
 \item[(ii)]
 for every strongly connected component ${\bf c}$ in the dependency graph of $\mp$ 
there is a def-module \hbox{${\text SM}_{\bf p}[F] \in \mp$} such that ${\bf p}$ contains 
all vertices in~${\bf c}$.
\end{enumerate}
It is easy to see, for example, that modular program~\eqref{eq:modprog1} is coherent. 
 
The following theorem is similar to the Splitting Theorem from \cite{fer09a}. 
That theorem says that under certain conditions the stable models of a 
conjunction of two formulas coincide with those interpretations that are stable 
models of both individual formulas with respect to different sets of intensional 
predicates. The theorem below presents a similar result for coherent programs and is more general in the sense that it applies to 
any finite number of \definitions, rather than just two.    

\begin{theorem}[Splitting Theorem]\label{thm:rsplit}
If $\mp$ is a coherent modular program 
then an interpretation $I$ is an  $\iota(\mp)$-stable model of 
%\beq
%\text{SM}_{\iota(\mp)}[
$\formula(\mp)$
%] 
%\eeq{eq:split}
iff it is a stable model of $\mp$.
\end{theorem}

Since modular program \eqref{eq:modprog1} is coherent, it is not by chance that its stable models coincide with the Herbrand
$p,q,r,s$-stable models of \eqref{eq:sm_prog}. Rather, this is an instance of 
a general fact. 
%For example, 
%first-order formula $S$ from Section \ref{sec:sm} is a rule. 
%It is easy to see 
%that disjunctive logic rules as introduced in~\citep{gel91b}  can be 
%represented using simple modular logic 
%programs. In fact, choice rules can also be represented in this syntax. For instance, we view 
%the choice rule ${\tt \{p(X)\}\ar q(X).}$ as an abbreviation for the rule $\forall x (q(X)\wedge \neg \neg p(x) \rar 
%p(x)).$
The following theorem, which follows from Theorems~\ref{thm:sm-traditional} and 
\ref{thm:rsplit},  describes the relationship between  modular programs and traditional logic programs. 
\begin{theorem}\label{thm:sm-traditional2}
For a coherent modular program $\mp$  such that $\sigma(\mp)$ contains at least 
one object constant and  $\pred(\mp)=\iota(\mp)$ 
%(in other words, the signature of $\mp$ coincides with the set of intensional 
%predicates of $\mp$) 
%if  $\sigma(\mp)$ contains at least one object constant then 
and any Herbrand interpretation $X$ of $\sigma(\mp)$
 the following conditions are equivalent
 \begin{itemize}
\setlength\itemsep{0em}
\item  $X$ is an answer set of $\formula(\mp)$;
\item $X$ is a stable model of $\mp$.   
\end{itemize}
\end{theorem}

A  modular program 
$
\{\hbox{SM$_{p}[q(1)\rar p(1)]$, SM$_{q}[p(1)\rar q(1)]$}\}
$
is an example of  a non-coherent program. 
Consider the Herbrand interpretation $\{p(1),q(1)\}$. This interpretation is a stable model of this program. 
Yet, it is not an answer set of the traditional program consisting of the two 
rules 
$q(1)\rar p(1).$ and
$p(1)\rar q(1).$
The only answer set of this traditional program 
is the empty set.

We now illustrate how modular programs capture the encoding~\eqref{eq:hc2} of the \hc so that each of its modules carries its intuitive meaning. 
The \hc modular program presented below consists of five \definitions: 
%\beq
% \ba{ll}
\begin{align}
&\text{SM}_{{\it edge}}[{\it edge}(a, a')\wedge \;\; \dots \;\; \wedge {\it 
edge}(c, c') \land 
\forall x y ({\it edge}(y, x) \rar {\it edge}(x,y)]\label{eq:hcmod1}\\
&\text{SM}_{{\it in}}[\forall x y( (\neg\neg  {\it in}(x,y) \wedge {\it edge}(x,y)) \rar {\it in}(x,y))]\label{eq:hcmod2}\\
&\text{SM}[\forall x y z(({\it in}(x,y)\wedge {\it in}(x,z)\wedge \neg(y=z) )\rar\bot) 
\wedge \label{eq:hcmod2.1}\\
&~~~~~~\forall x y z(({\it in}(x,z)\wedge {\it in}(y,z)\wedge \neg(x=y)) 
\rar\bot) \wedge
\nonumber\\
&~~~~~~\forall x y (({\it in}(x,y)\wedge {\it in}(y,x)) 
\rar\bot)] 
\nonumber\\
&\text{SM}_{{\it r}}[\forall x y({\it in}(x,y) \rar {\it r}(x, y)) \wedge\label{eq:hcmod3}\\
&~~~~~~~\forall x y z(({\it r}(x,z) \wedge {\it r}(z,y))\rar {\it r}(x,y))] \nonumber
\\
&\text{SM}[\forall x y z z'((\neg {\it r}(x, y)\wedge {\it edge}(x, z)\wedge 
{\it edge}(z', y))\rar \bot)]\label{eq:hcmod4}
\end{align}
We call this  modular program $\mh$. 
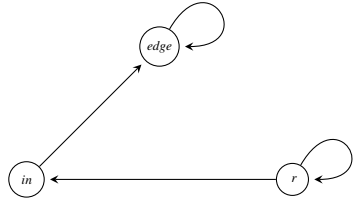
\begin{figure}
\centering
\tiny
\begin{tikzpicture}[shorten >=2pt, auto]
  \node[place] (e) {{\it edge}};
  \node[place] (i) [below left of=e] {$\;\;${\it in}$\;\;$ };
  \node[place] (r) [below right of=e] {$\;\;\;${\it r}$\;\;$ };
  \path[->]
     (r) edge [loop above] node {} ()
         edge[->] node[above] {} (i)
     (i) edge[->] node[below] {} (e)
     (e) edge [loop above] node {} ();
\end{tikzpicture}
\normalsize
\caption{Dependency graph for $\mh$.} 
\label{fig:dpghc}
\end{figure}
%{\color{blue} Amelia: below I use notation predicate symbol superscript interpretation as explained in lecture notes on mathematical logic by Vladimir. We probably should define that. I also wonder if I am too wordy at times and certain things could be explained shorter using some terminology or ways of expressing concepts from logic. In particular, see narrative about \definition~\eqref{eq:hcmod1}. What do you think?}
The \definitions shown above 
correspond to the intuitive groupings of rules of the \hc encoding discussed in Section~\ref{sec:trpr}. 
\begin{itemize} 
\setlength\itemsep{0em}
\item  An {\it edge}-stable model of \definition~\eqref{eq:hcmod1} is any 
interpretation $I$ over $\sigma(\mh)$ such that the extension\footnote{The {\sl extension} of a predicate in an interpretation is the set of 
tuples that satisfy the predicate in that interpretation.} of the ${\it 
edge}$ predicate in $I$ corresponds to the symmetric closure of the facts in~\eqref{eq:hcfacts}. 
%(i) it satisfies the conjunction
%\beq
%{\it edge}(a, a')\wedge \;\; \dots \;\; \wedge {\it edge}(c, c')
%\eeq{eq:edgedef1}
%and (ii) 
%constructing an interpretation $I'$ from $I$ so that it differs from $I$ only in interpreting {\it edge} 
%as follows ${\it edge}^{I'}\subseteq{\it edge}^I$ always results in $I'$ that does not satisfy~\eqref{eq:edgedef1}.
 \item An {\it in}-stable model of  \definition~\eqref{eq:hcmod2} 
is any interpretation $I$ over $\sigma(\mh)$ such that 
the extension of the predicate ${\it in}$ in $I$ is a subset of the extension of 
the predicate ${\it edge}$ in $I$. 
%${\it in}^I\subseteq{\it edge}^I.
\item An $\emptyset$-stable model of  \definition~\eqref{eq:hcmod2.1} 
is any interpretation $I$ over $\sigma(\mh)$ that satisfies the conjunction in~\eqref{eq:hcmod2.1}.
\item An ${\it r}$-stable model of  \definition~\eqref{eq:hcmod3} 
is any interpretation $I$ over $\sigma(\mh)$, where relation {\it r} is the transitive closure of relation {\it in}.
\item An $\emptyset$-stable model of  \definition~\eqref{eq:hcmod4} 
is any interpretation $I$ over $\sigma(\mh)$ that satisfies the conjunction in~\eqref{eq:hcmod4}.
\end{itemize}
Any interpretation over $\sigma(\mh)$ that satisfies the conditions imposed by 
every individual module of $\mh$ is a stable model of $\mh$. 

The dependency graph of $\mh$ is shown in Figure \ref{fig:dpghc}. The strongly 
connected components of this graph each consist of a single vertex. It is easy 
to verify that   
the \hc  program $\mh$ is coherent. 
%The union of its components results in~\eqref{eq:hc2}.
By Theorem~\ref{thm:sm-traditional2}, it follows that the Herbrand models of \hc coincide with the answer sets of~\eqref{eq:hc2} so that answer set solvers can be used to find these models. 

Arguably, when answer set practitioners develop their applications they 
intuitively associate meaning with components of the program. We believe that 
modular programs as introduced here provide us with a suitable model for 
understanding the meaning of components of the program.

\section{Conservative Extensions}%and Equivalent Rewritings}
\label{sec:eqre}
In this section, we study the question of how to formalize common rewriting techniques used in answer set programming, such as projection, and argue their correctness. 

Let $F$ and $G$ be  second-order formulas such that
$\pred(F)\subseteq\pred(G)$ and both formulas share the same function symbols. 
We say that~{\em $G$ is a conservative extension of $F$} if 
%their models are in 1-1 correspondence so that
%$$ \{M \mid M \text{ is a model of } F\}
%=
% \{M_{|_{\sigma(F)}} \mid M \text{ is a model of } G\}.$$
%\begin{enumerate}
%$M \mapsto M|_{\sigma(F)}$ is a 1-1 correspondence between the models of $G$ and 
%the models of $F$. 
\begin{itemize}
\item\label{cond:eeq1} $\{M | M \text{ is a model of } F\}
=
 \{M_{|\sigma(F)} | M \text{ is a model of } G\},$ and
\item\label{cond:eeq2} there are no distinct models $M$ and $M'$ of $G$ such 
that $M_{|\sigma(F)} = M'_{|\sigma(F)}$. 
\end{itemize}
The definition of a conservative extension for second-order formulas gives us 
a definition of a conservative extension for \definitions, as they are second-order formulas.
It is interesting to note that the first condition
of the definition holds 
if and only if $F$ 
has the same models as the second-order formula $\exists p_1 \dots p_n \; G,$
where $\{p_1, \dots p_n\}=\pred(G)\setminus\pred(F).$
The second condition %~\ref{cond:eeq2} 
adds another intuitive restriction. For example, consider the broadly used 
Tseitin transformation. In this transformation, an 
arbitrary propositional formula is converted into conjunctive normal form by (i) 
augmenting the original formula with ``explicit definitions'' and (ii) applying 
equivalent transformations. The resulting formula is of a new signature, but 
both of the 
conditions %~\ref{cond:eeq1} and~\ref{cond:eeq2} 
of the definition hold between the original formula and the result of Tseitin transformation. 
We can state the definition of a conservative 
extension more concisely by saying that $G$ is  
a conservative extension of $F$ if $M \mapsto M_{|\sigma(F)}$ is a 1-1 
correspondence between the models of $G$ and the models of $F$. 

In view of Theorem~\ref{thm:sm-traditional}, the definition of a conservative 
extension can be applied to traditional logic programs:
If $\Pi_1$ and $\Pi_2$ are  traditional  programs such that
$\pred(\Pi_1)\subseteq\pred(\Pi_2)$ and both programs share the same function symbols, 
then  $\Pi_2$ is a conservative extension of $\Pi_1$ if \hbox{$M \mapsto M_{|\sigma(\Pi_1)}$} is a 1-1 correspondence between the answer sets of $\Pi_2$ and 
the answer sets of~$\Pi_1$. 

%\noindent 
As an illustration of a conservative extension, consider the following formulas:
\begin{align}
& \forall xz(s(x,z)\lrar (p(z) \wedge \exists y(q(x,y) \wedge r(x,y)))) 
\label{eq:sms1}\\
& \forall xz(s(x,z)\lrar (p(z) \wedge t(x))) \wedge
  \forall v(t(v)\lrar\exists w(q(v,w) \wedge r(v,w))).
\label{eq:sms2t}
\end{align}

%\begin{align}
%& \forall xyz(s(x,z)\lrar (p(z) \wedge q(x,y) \wedge r(x,y)))
%\label{eq:sms1} \\
%&  \forall xz(s(x,z)\lrar  (p(z) \wedge t(x))) \wedge
%  \forall xy(t(x)\lrar(q(x,y) \wedge r(x,y))).
%\label{eq:sms2t}
%\end{align}
It is easy to 
verify that the models of formulas~\eqref{eq:sms1} and~\eqref{eq:sms2t}  are in 1-1 correspondence so that
$$
\ba{c}
 \{M \mid M \text{ is a model of formula~\eqref{eq:sms1}}\}
=
\{M_{|\{s,p,q,r\}} \mid M \text{ is a model of  formula~\eqref{eq:sms2t}}\}.
\ea
$$
%a interpretation $M$ is a model of \eqref{eq:sms1} iff there is an 
%extension of $M$ to the signature $\sigma \cup \sigma'$ that is 
%a model of \eqref{eq:sms2t}.
In fact, formula~\eqref{eq:sms2t} is obtained from formula~\eqref{eq:sms1} by 
introducing an explicit definition using predicate symbol $t$. % and applying  equivalent 
% transformations from classical logic. 
Recall the notion of an explicit definition:
to extend a formula $F$ by {\em an explicit definition using predicate symbol 
$t$} 
means to add to the signature of $F$ a new predicate symbol $t$ of arity $n$, 
and to add a conjunctive term to~$F$ of the form 
\beq
\forall x_1\dots x_n (t(x_1,\dots,x_n)\leftrightarrow G),
\eeq{eq:expldef}
where $x_1\dots x_n$ are distinct variables and $G$ is a formula over the signature of $F$.
The result of adding such a definition is a formula that is a conservative extension of $F$.
Furthermore, constructing a formula from $F$ by 
\begin{itemize}
 \item substituting every occurrence of subformula $G$ in $F$  with 
$t(x_1,\dots,x_n)$ (modulo proper substitution of terms) and
 \item extending this formula with a conjunctive term~\eqref{eq:expldef}
\end{itemize}
results in a conservative extension as well. This is the procedure that is used 
to obtain formula~\eqref{eq:sms2t} from~\eqref{eq:sms1}. 
%An example of extending \eqref{eq:sms1} with an explicit definition using predicate symbol $t$ 
%follows:
%For example, the following formula is an example of extending \eqref{eq:sms1} 
%with an explicit definition using predicate symbol $t$:  
%from~\eqref{eq:sms1} using this procedure. 
%\beq
%$$\forall xz(s(x,z)\lrar (p(z) \wedge \exists y(q(x,y) \wedge r(x,y)))) \wedge
%  \forall v(t(v)\lrar\exists w(q(v,w) \wedge r(v,w))).
%$$
%\eeq{eq:expldefex}
%\smallskip
%\beq
%\forall xz(s(x,z)\lrar (p(z) \wedge t(x))) \wedge
%  \forall v(t(v)\lrar\exists w(q(v,w) \wedge r(v,w))).
%\eeq{eq:expldefex.2}
%This formula is constructed %and is  
%equivalent to~\eqref{eq:expldefex}. It is easy to see that 
%formula~\eqref{eq:expldefex.2} is equivalent to 
%to formula~\eqref{eq:sms2t}.

Recall that $S$ denotes sentence~\eqref{eq:s1def}.
By $S'$ we denote the sentence
\beq
\forall x z ((t(x) \wedge p(z)) \rar s(x,z)) 
\land 
\forall x y ((q(x,y) \wedge r(x,y)) \rar t(x)).
\eeq{eq:sprime1def}
It can be verified that \eqref{eq:sms1} is equivalent to 
SM$_{s}[S], $ and that \eqref{eq:sms2t} is equivalent to 
${\text SM}_{s,t}[S'].
$

The next proposition provides a general method for showing 
that one \definition is a conservative extension of another.
%essential equivalence
%between \definitions.

\begin{prop}\label{obs:eseqdef}
For any \definitions SM$_{\bf p}[F]$ and SM$_{\bf p \cup \bf p'}[G]$ such that
$\pred(F)\subseteq\pred(G)$, both formulas share the same function symbols,  and 
${\bf p}'$  is a subset of predicate symbols 
 $\pred(G)\setminus\pred(F)$,
if SM$_{\bf p}[F]$ and SM$_{\bf p \cup \bf p'}[G]$ are equivalent to first-order 
formulas $F'$ and $G'$ respectively, and $G'$ is a conservative extension of  $F'$ 
%are essentially equivalent, 
then
SM$_{\bf p \cup \bf p'}[G]$ is a conservative extension of
\smpf.  
%are essentially equivalent \definitions. 
\end{prop}

%The same reasoning also shows that 
An analogous property holds for traditional programs:
\begin{prop}\label{obs:eseqdef2}
 For any traditional programs $\Pi_1$ and $\Pi_2$
  such that 
$\pred(\Pi_1)\subseteq\pred(\Pi_2)$ and both programs share the same function symbols and contain at least one object constant,  
if SM$_{\pred(\Pi_1)}[\Pi_1]$ and SM$_{\pred(\Pi_2)}[\Pi_2]$ are equivalent to first-order 
formulas $\Pi_1'$ and $\Pi_2'$ respectively, and $\Pi_2'$ is a conservative 
extension of $\Pi_1'$, 
%are essentially equivalent, 
then
traditional program  $\Pi_2$ is a conservative extension of  $\Pi_1$.
% are essentially equivalent.
\end{prop}

We now lift the definition of a conservative extension to the case of modular programs. 
 We say that modular 
program {\em $\mp'$ is a conservative extension of $\mp$} 
%are {\em essentially equivalent} 
if $M \mapsto M_{|\sigma(\mp)}$ is a 1-1 correspondence between the models of 
$\mp'$ and 
the models of $\mp$. 
%\begin{itemize}
%\item $\{M | M \text{ is a stable model of } \mp\}
%=
% \{M_{|\sigma(\mp)} | M \text{ is a stable model of } \mp'\},$ and
%\item there is a 1-1 correspondence between the models of $\mp$ and $\mp'$. 
%\end{itemize}

Let us recall the notion of strong equivalence~\citep{lif01}. 
Traditional programs $\Pi_1$ and~$\Pi_2$ are {\em strongly equivalent} if for 
every traditional program $\Pi$, programs $\Pi_1\cup \Pi$ and $\Pi_2\cup \Pi$ have the same answer sets.
Strong equivalence %is a powerful property and 
can be used to argue the 
correctness of some program rewritings used by answer set programming 
practitioners. However, the projection rewriting technique, 
exemplified by replacing rule~\eqref{eq:s1def} with rules~\eqref{eq:sprime1def}, 
cannot be justified using the notion of strong equivalence. This rewriting 
technique is commonly used to improve the performance of answer set 
programs~\citep{bud15}. Strong equivalence is inappropriate  for 
justifying this rewriting for a simple reason: the signature of the original 
program is changed. In what follows we attempt to ``adjust'' the notion of strong 
equivalence to the context of modular programs so that we may formally reason about 
the correctness of 
projection and other similar rewriting techniques. We then translate these notions to the realm of traditional programs.
We start by restating the definition of strong equivalence given in~\citep{fer09} and recalling some of its properties.
 
First-order formulas $F$ and $G$ are {\em strongly equivalent} if 
for any formula $H$, any occurrence of~$F$ in~$H$, and any list~{$\bf p$} of 
distinct predicate constants,  SM$_{\bf p}[H]$ is equivalent to 
SM$_{\bf p}[H']$, where $H'$ is obtained from $H$ by replacing 
$F$ by $G$. In~\citep{lif07a} the authors show that first-order 
formulas $F$ and~$G$ are strongly equivalent if they are intuitionistically equivalent.

The following theorem, which is easy to verify, illustrates that classical equivalence between 
second-order formulas is sufficient to capture the condition of ``strong equivalence'' for modular programs.
In other words, replacing a \definition  
by an equivalent \definition with the same intensional predicates does not change 
the semantics of a modular program.
\begin{theorem}\label{thm:streq1} 
Let SM$_{\bf p}[F]$ and SM$_{\bf p}[G]$ be  \definitions.
Then the following two conditions are equivalent:
\begin{enumerate}
 \item[(i)] for any modular program $\mp$, the programs $\mp\cup\{ 
\hbox{SM$_{\bf p}[F]$} \}$
and $\mp\cup\{ \hbox{SM$_{\bf p}[G]$} \}$ have the same stable models;
 \item[(ii)] SM$_{\bf p}[F]$ 
and SM$_{\bf p}[G]$ are equivalent.
\end{enumerate}
\end{theorem}
In~\citep[Section~5.2]{fer09}, the authors observe that
if first-order formulas 
$F$ and $G$ are strongly equivalent then \definitions of the form SM$_{\bf 
p}[F]$ and SM$_{\bf p}[G]$ are equivalent. Consequently, to show that 
replacing SM$_{\bf p}[F]$ by SM$_{\bf p}[G]$ in any modular program results in 
a program with the same models it is sufficient to prove that $F$ and 
$G$ are intuitionistically equivalent. 

The following theorem  lifts Theorem~\ref{thm:streq1} to conservative extensions.

%\begin{theorem}\label{thm:SMstrongeq}
%Let $F$,$G$ be traditional programs, and let
%{\bf p} be a tuple constructed from any predicates occurring in $F$ or $G$.  
%\end{theorem}

%We say that \definition SM$_{\bf p}[F]$ is {\em essentially strongly equivalent} 
%to \definition SM$_{{\bf p}\cup{\bf p}'}[G]$ when for any modular program $\mp$ 
%such that $\pred(\mp)$ contains no elements from ${\bf p}'$, modular programs $\mp\cup\{ \hbox{SM$_{\bf p}[F]$} \}$
%and $\mp\cup\{ \hbox{SM$_{{\bf p}\cup {\bf p}'}[G]$} \}$ are essentially equivalent.

\begin{theorem}\label{thm:streq2} 
Let SM$_{\bf p}[F]$, SM$_{{\bf p}\cup{\bf p}'}[G]$ be  \definitions such that 
$\pred(F)\subseteq\pred(G)$, both formulas share the same function symbols,  and 
${\bf p}'$  is %a subset of predicate symbols 
 $\pred(G)\setminus\pred(F)$.
Then the following two conditions are equivalent:
\begin{enumerate}
 \item[(i)] for any modular program $\mp$ 
such that $\pred(\mp)$ contains no elements from ${\bf p}'$, 
modular program $\mp\cup\{ \hbox{SM$_{{\bf p}\cup {\bf p}'}[G]$} \}$ 
is a conservative extension of 
$\mp~\cup~\{ \hbox{SM$_{\bf p}[F]$} \}$;
 
%are essentially equivalent
 \item[(ii)] SM$_{{\bf p}\cup {\bf p}'}[G]$ is a conservative extension of 
SM$_{\bf p}[F]$. 
%are essentially equivalent.
\end{enumerate}
\end{theorem}
Theorem~\ref{thm:streq2} tells us that we can replace \definitions in a modular 
program with their conservative extensions and are guaranteed to obtain 
a conservative extension of the original modular program.  Thus, conservative 
extensions of 
\definitions allow us to establish something similar to strong equivalence for 
modular programs with possibly different signatures. %traditional answer set programming.

For example, consider the choice rule $\tt{\{p\}}$, a shorthand for the rule 
%\beq
$p \ar \no \no p$. In some answer set programming dialects double negation is not allowed in the body of a rule.
It is then common to simulate a choice rule as above by introducing an auxiliary atom $\hat{p}$ and using the rules  $\neg \hat{p} \rar p$ and $\neg p \rar \hat{p}$. It is easy to check that 
SM$_{p,\hat{p}}[(\neg \hat{p}\rar p)\land(\neg p\rar\hat{p})]$ is a conservative extension of 
SM$_p[p \lor \neg p]$. By Theorem~\ref{thm:streq2}, it follows that we can replace the latter 
with the former within the context of any modular program not containing the 
predicate symbol $\hat{p}$, and get a conservative extension of the original program.

Proposition~\ref{obs:eseqdef} and Theorem~\ref{thm:streq2} equip us with a method for 
establishing the correctness of program rewritings.
For instance, the fact that formulas \eqref{eq:sms1} and~\eqref{eq:sms2t}
are equivalent to \definitions
SM$_{s}[S]$ and  
${\text SM}_{s,t}[S']$ respectively, translates into the assertion that 
the latter is a conservative extension of the former. 
%se \definitions are essentially equivalent.
Thus, replacing \definition SM$_{s}[S]$ in modular program~\eqref{eq:modprog1} 
with ${\text SM}_{s,t}[S']$ results in a modular program that is a conservative 
extension of~\eqref{eq:modprog1}.
Similarly, replacing \definition~\eqref{eq:hcmod4} in the \hc modular program 
presented in Section~\ref{sec:relprog} by the \definition
\beq
\ba{l}
\text{SM}_{vertex1,vertex2}[
\forall x y ((\neg {\it r}(x, y)\wedge {\it vertex1}(x) \land {\it vertex2}(y) 
\rar \bot) \;\; \wedge \\
\forall x z({\it edge}(x,z) \rar {\it vertex1}(x)) \;\;\land \\
\forall z'y({\it edge}(z',y) \rar {\it vertex2}(y))]
\ea
\eeq{eq:hcmodpr4}
results in a conservative extension of the original program.
% \definitions~\eqref{eq:hcmod4} and~\eqref{eq:hcmodpr4} are essentially  equivalent.  
This is an instance of projection rewriting. We now introduce some notation used to state a result about the general case of projection that will support our claim that 
\eqref{eq:hcmodpr4} is a conservative extension of~\eqref{eq:hcmod4}. 
%are essentially equivalent.  

Let~$R$ be a rule~\eqref{eq:rule} occurring in a
 traditional logic program $F$, and let 
{\bf x} be a non-empty tuple of variables occurring only in 
%atoms in 
%the positive part of the 
the body of~$R$. By $\alpha({\bf x,y})$ we denote the conjunction of all conjunctive terms in the body 
of~$R$ that contain at least one variable from {\bf x}, where {\bf y} denotes 
all the variables occurring in these conjunctive terms but not occurring in {\bf x}. 
By $\beta$ we denote the set of all conjunctive terms in the body of~$R$ that 
do not contain any variables occurring in ${\bf x}$. By $\gamma$ we denote the head of~$R$. Let $t$ be a 
predicate symbol that does not occur in $F$. Then the 
\emph{result of projecting variables~${\bf x}$ out of~$R$ using predicate symbol $t$} is the conjunction of the 
following two rules  
%\beq
\begin{align*}
\widetilde{\forall}\left ( \left ( t({\bf y}) \land \beta \right ) \rar \gamma 
\right ),\\
%\eeq{eq:r1x}
%\beq
{\forall {\bf x y}}\left ( \alpha({\bf x,y}) \rar 
t({\bf y})\right ).
\end{align*}
%\eeq{eq:r2x}
For example, the result of projecting $y$ out of \eqref{eq:s1def} using predicate symbol $t$ is 
\eqref{eq:sprime1def}. 
We can project variables out of a 
traditional logic program by successively projecting variables out of rules. 
For example, first projecting 
$z$ out of the traditional logic program in \eqref{eq:hcmod4} and then 
projecting $z'$ out of the first rule of the resulting program yields  
the traditional logic program in~\eqref{eq:hcmodpr4}. 

%This definition of projection can be intuitively extended to traditional
%formulas that go beyond single rule so that we can talk about projecting 
%variables out of a  \definition. 

\begin{theorem}\label{thm:projection} Let SM$_{p_1, \dots, p_k}[F]$ be a 
\definition and~$R$ be a rule in $F$. Let $\bx$ denote
a non-empty tuple of variables occurring in the body of~$R$, but 
not in the head.
If~$G$ is constructed from $F$ by replacing~$R$ in~$F$ with the result of 
projecting variables ${\bf x}$ out of~$R$ using a predicate symbol~$p_{k+1}$ 
that is not in the signature of $F$,
then SM$_{p_1, \dots, p_{k+1}}[G]$ 
is a conservative extension of 
SM$_{p_1, \dots, p_k}[F]$. 
%and are essentially equivalent.
\end{theorem}

We now restate Theorem~\ref{thm:projection} in terms of traditional logic programs using the link between \definitions and traditional programs established in Theorem~\ref{thm:sm-traditional}.
\begin{cor}\label{cor:projection} Let $\Pi$ be a traditional logic program containing at least one object constant
and~$R$ be a rule in~$\Pi$. Let $\bx$ denote
a non-empty tuple of variables occurring in the body of~$R$, but 
not in the head.
If $\Pi'$ is constructed from $\Pi$ by replacing~$R$ in $\Pi$ with the result of 
projecting variables~${\bf x}$ out of~$R$ using a predicate symbol $p$ that  
does not occur in $\Pi$,
then $\Pi'$ 
is a conservative extension of $\Pi$. 
%are essentially equivalent.
\end{cor}

\section{Conclusion}
In this paper, we introduced first-order modular logic programs that provide a 
way of viewing logic programs as consisting of many independent, meaningful 
modules. We also defined conservative extensions, which like 
strong equivalence for traditional programs, can be useful for reasoning 
about traditional programs and modular programs. We showed how these concepts 
may be used to justify the 
common projection rewriting.  

\section*{Acknowledgments}
Many thanks to Joshua Irvin, Vladimir Lifschitz, and Miroslaw Truszczynski for useful discussions regarding ideas in this 
paper. Thanks as well to the anonymous referees for helpful comments. Amelia Harrison was partially supported by the
National Science Foundation under Grant IIS-1422455.  

\bibliography{bib}

\newpage
\appendix
\section{Appendix: Proofs of Theorems}

\subsection{Proof of Splitting Theorem (Theorem~\ref{thm:rsplit})} 
\noindent {\bf Splitting Theorem.}\emph{
		If $\mp$ is a coherent modular program 
		then an interpretation $I$ is an  $\iota(\mp)$-stable model of 
		%\beq
		%\text{SM}_{\iota(\mp)}[
		$\formula(\mp)$
		%] 
		%\eeq{eq:split}
		iff it is a stable model of $\mp$.
	}
\medskip

\begin{proof}
Let $\mp$ be $\{{\text SM}_{\bf p_1}[F_1], \dots , 
{\text SM}_{\bf p_n}[F_n]\}$. 
The proof is by induction on $n$. The base case is trivial. In the induction 
step, we assume that
for any simple modular program $\mp$ of the form  $$\{{\text SM}_{\bf p_1}[F_1], \dots , 
{\text SM}_{\bf p_k}[F_k]\}$$ and  meeting conditions (i) and (ii) of a coherent program, $I$ is a stable 
model of $\mp$ 
iff it is an
$\iota(\mp)$-stable model of 
$\formula(\mp)$.
Consider a simple modular program $$\mp' = \{{\text SM}_{\bf p_1}[F_1], \dots , 
{\text SM}_{\bf p_k}[F_k], {\text SM}_{\bf p_{k+1}}[F_{k+1}]\}$$ meeting 
conditions (i) and~(ii). Let $\mp'_k \subset \mp'$ denote the set 
$\{{\text SM}_{\bf p_1}[F_1], \dots , 
{\text SM}_{\bf p_k}[F_k]\}$.   
Now, an interpretation $I$ is an 
$\iota(\mp')$-stable model of
$
%\text{SM}_{\iota(\mp')}[
\formula(\mp')%] 
$
iff it satisfies the formula 
$$
\text{SM}_{\iota(\mp')}[\bigwedge_{1 \leq i \leq k}F_i \;\; \land \;\; F_{k+1}].  
$$
But by the Splitting Theorem from \citep{fer09a}, this is the case iff 
$I$ satisfies  
\beq
\text{SM}_{\iota(\mp'_k)}[\bigwedge_{1 \leq i \leq k}F_i] \;\; \land \;\; 
\text{SM}_{\iota(F_{k+1})}[F_{k+1}],  
\eeq{eq:splitn+1}
which is true iff 
$I$ satisfies both conjunctive terms.  
But $I$ satisfies 
$$
\text{SM}_{\iota(\mp'_k)}[\bigwedge_{1 \leq i \leq k}F_i]
$$
iff it is an 
$\iota({\mp'_k})$-stable model of 
$\formula(\mp)$, and by the 
induction hypothesis, this 
is the case iff $I$ is a stable model of $\mp'_k$. 
Interpretation $I$ is a stable model of $\mp'_k$  
iff it satisfies SM$_{p_i}[F_i]$ for $1 \leq i \leq k$. 
So $I$ satisfies~\eqref{eq:splitn+1} iff it satisfies 
SM$_{p_i}[F_i]$ for $1 \leq i \leq k+1$, 
which is the case iff $I$ is a stable model 
of $\mp'$. 
\end{proof}

\subsection{Proofs of Propositions~\ref{obs:eseqdef} and \ref{obs:eseqdef2}} 
\noindent{\bf Proposition \ref{obs:eseqdef}.}\emph{
For any \definitions SM$_{\bf p}[F]$ and SM$_{\bf p \cup \bf p'}[G]$ such that
$\pred(F)\subseteq\pred(G)$, both formulas share the same function symbols, and 
${\bf p}'$ is a subset of predicate symbols 
 $\pred(G)\setminus\pred(F)$,
if SM$_{\bf p}[F]$ and SM$_{\bf p \cup \bf p'}[G]$ are equivalent to first-order 
formulas $F'$ and $G'$ respectively, and $G'$ 
is a conservative extension of $F'$,
then
SM$_{\bf p \cup \bf p'}[G]$ 
is a conservative extension of 
SM$_{\bf p}[F]$. 
} 

\medskip

\begin{proof}
Consider \definitions \smpf and SM$_{\bf p \cup \bf p'}[G]$ and first-order 
formulas $F'$ and $G'$, meeting the conditions of the 
proposition. Then first-order formula $F'$ has the same models as \smpf, and 
first-order formula $G'$ has the same models as  
SM$_{\bf p \cup \bf p'}[G]$. Furthermore, since $G'$ is a conservative extension 
of $F'$, $M \mapsto M_{|\sigma(F')}$ is a 1-1 correspondence between the models 
of $G'$ and the models of $F'$. It follows that this function is also a 1-1 
correspondence between the models of \smpf and SM$_{\bf p \cup \bf p'}[G]$. 
\end{proof}

The same reasoning  shows that Proposition~\ref{obs:eseqdef2} holds

\begin{comment}
\end{comment}

\begin{comment}
{\color{blue} Yulia: It seems to me that theorem 6 now is the same as the 
proposition above, and the theorem below no longer occurs in the note. Is this 
what we want?}

\noindent {\bf Theorem \ref{thm:SMstrongeq}.} 
\emph{Let $F$,$G$ be traditional programs, and let
{\bf p} be a tuple constructed from any predicates occurring in $F$ or $G$.  If 
$F$ and $G$ are strongly equivalent then \definitions SM$_{\bf p}[F]$ and 
SM$_{\bf p}[G]$ are strongly equivalent.}

\medskip

\begin{proof}
Consider two strongly equivalent formulas $F$ and 
$G$. Since they are strongly equivalent, $\pi(F) = \pi(G)$. Let $\bp$ be a 
subset of $\pi(F)$. By ${\bf q}$ we'll denote the set $\pi(F) \setminus \bp$. 
For any formula $q$, let $\choice{q}$ denote the formula $\forall x(q(x) 
\lor \neg q(x))$, and let $\choice{{\bf q}}$ denote the conjunction of the 
formulas $\choice{q}$ for all $q$ in ${\bf q}$.
Since $F$ and $G$ are strongly equivalent, $F \land \choice{{\bf q}}$ has the 
same stable models as $G \land \choice{{\bf q}}$. By Proposition 
\ref{prop:strongclass}, it follows that 
SM$_{\bp{\bf q}}[F \land \choice{{\bf q}}]$
is equivalent to 
SM$_{\bp{\bf q}}[G \land \choice{{\bf q}}]$. But by Theorem~2 from \cite{fer09},  
SM$_{\bp{\bf q}}[F \land \choice{{\bf q}}]$ is equivalent to 
\smpf, and  
SM$_{\bp{\bf q}}[G \land \choice{{\bf q}}]$ is equivalent to 
SM$_{\bf p}[G]$, so that \smpf is equivalent to  
SM$_{\bf p}[G]$. 
\end{proof}
\end{comment}
\subsection{Proof of Theorem \ref{thm:streq2}}  

\noindent {\bf Theorem \ref{thm:streq2}.} \emph{ 
Let SM$_{\bf p}[F]$, SM$_{{\bf p}\cup{\bf p}'}[G]$ be  \definitions such that
$\pred(F)\subseteq\pred(G)$, both formulas share the same function symbols,  and 
${\bf p}'$  is %a subset of predicate symbols 
 $\pred(G)\setminus\pred(F)$,
then the following two conditions are equivalent
\begin{enumerate}
 \item[(i)] for any modular program $\mp$ 
such that $\pred(\mp)$ contains no elements from ${\bf p}'$, 
modular programs $\mp\cup\{ \hbox{SM$_{{\bf p}\cup {\bf p}'}[G]$} \}$
is a conservative extension of 
$\mp~\cup~\{\smpf\}$.
 \item[(ii)] SM$_{{\bf p}\cup {\bf p}'}[G]$ 
is a conservative extension of SM$_{\bf p}[F]$.
\end{enumerate}}
\begin{proof}
Establishing that if condition (i) holds then condition (ii) also holds is not 
difficult. In the other direction, assume 
SM$_{{\bf p}\cup {\bf p}'}[G]$  
is a conservative extension of \smpf. %Then $\pi(F) \subseteq \pi(G)$  and an interpretation $M$ is a model of 
%\smpf  iff there is 
%some $M'$ over the signature $\sigma(G)$ such that $M'_{|\sigma(F)} = M$ and $M'$ is a  model of 
%SM$_{{\bf p}\cup {\bf p}'}[G]$.  
%Now consider an arbitrary modular program 
We need to show that for any modular program $\mp$ such that $\pi(\mp)$ does not 
contain any elements of $\bp'$,  $\mp \cup 
\{\text{SM}_{{\bf p}\cup {\bf p}'}[G]\}$ 
is a conservative extension of 
$\mp \cup \{\smpf\}$. 
%are essentially equivalent.
Let $M$ be 
%an interpretation over $\sigma(\mp)\cup \sigma(F)$. This 
%interpretation is a stable model of  
a model of $\mp \cup\{ \smpf\}$.  Then %iff  
\begin{enumerate}
\item [(a)] $M_{|\sigma(F)}$ is a 
model of \smpf and 
\item [(b)] $M_{|\sigma(H)}$ is a model    
 of each \definition SM$_{{\bf q}}[H]$ in $\mp$. 
\end{enumerate}
By our initial assumption, %condition (a) holds of $I$  
%iff there is  
%some 
$M_{|\sigma(F)}$ can be extended to the signature $\sigma(G)$.
That is, there is some $M'$
 such that $M'_{|\sigma(F)} = 
M_{|\sigma(F)}$ and $M'$ is a  model of 
SM$_{{\bf p}\cup {\bf p}'}[G]$. 
Furthermore, there is a unique $M'$ about which the above property holds (recall the condition on 1-1 correspondence).
Since the signature of $G$ differs from 
the signature of $F$ only by predicates in ${\bf p}'$, and that none of these 
predicates occur in the signature of $\mp$,  
$M_{|\sigma(\mp)} \cup M'$ is an interpretation over $\sigma(\mp)\cup 
\sigma(G)$. Furthermore, it is clear that this interpretation is a model of
$\mp \cup \{\text{SM}_{{\bf p}\cup {\bf p}'}[G]\}$. Finally, it is easy to show that 
if $M$ is a  model of $\mp \cup \{\text{SM}_{{\bf p}\cup {\bf p}'}[G]\}$ 
then $M_{|\sigma(\mp) \cup \sigma(F)}$ is a model of $\mp \cup \{\smpf\}$. 
From the uniqueness of $M'$ the 1-1 correspondence condition of the definition 
of conservative extensions for modular programs also holds.
%We denote this interpretation by $N$.
%From (a) and our assumption it immediately follows that $N_{|\sigma(G)}$ is a model of SM$_{{\bf p}\cup {\bf p}'}[G]$. Let SM$_{{\bf t}}[H]$ be any \definition in $\mp$. It is easy to see that
%$N_{|\sigma(H)}=M_{|\sigma(H)}$.  By (b) it follows that $N$ is a model of SM$_{{\bf t}}[H]$.
%Consequently, $N$ is a stable model of SM$_{{\bf p}\cup {\bf p}'}[G]$ so that $N_{|\sigma(\mp)}=M$. By definition, SM$_{\bf p}[F]$ 
%and SM$_{{\bf p}\cup {\bf p}'}[G]$ are essentially equivalent.
\end{proof}

\subsection{Proof of Theorem~\ref{thm:projection}}

\noindent {\bf Theorem~\ref{thm:projection}.} \emph{Let 
SM$_{p_1, \dots, p_k}[F]$ be a \definition and 
$R$ be a rule in $F$ so that $\bx$ denotes
a non-empty tuple of variables occurring in atoms in the body of~$R$, but 
not in the head.
Let formula $G$ be constructed from $F$ by replacing~$R$ in $F$ with the result of 
projecting variables ${\bf x}$ out of~$R$ using predicate symbol $p_{k+1}$ not in the signature of $F$.
Then SM$_{p_1, \dots, p_{k+1}}[G]$ 
is a conservative extension of 
SM$_{p_1, \dots, p_k}[F]$. 
%and are essentially equivalent.
}
\begin{proof}
By the definition of projection, formula $G$ is constructed from  $F$ by replacing rule~$R$ in $F$ of the form~\eqref{eq:rule} with rules 
\beq
\widetilde{\forall}\left ( \left ( p_{k+1}({\bf y}) \land \beta \right ) \rar \gamma \right ),
\eeq{eq:r1x}
and
\beq
{\forall {\bf x y}}\left ( \alpha({\bf x,y}) \rar 
p_{k+1}({\bf y})\right ),
\eeq{eq:r2x}
where we assume the notation introduced in the end of Section~\ref{sec:eqre}.
%Recall that the variables ${\bf x}$ are being projected.
Consider minimizing the scope of the quantifiers in rule~$R$ as follows
\beq
\widetilde{\forall}\left ( \left ( (\exists {\bf x}\; \alpha({\bf x,y})) \land \beta \right ) \rar \gamma \right ).
\eeq{eq:r1x.2}
The transformation from~$R$ to~\eqref{eq:r1x.2} is an intuitionistically equivalent transformation. 
Thus~$R$ and~\eqref{eq:r1x.2} are strongly equivalent formulas.
Let $F'$ denote the result of replacing~$R$ in $F$ by \eqref{eq:r1x.2}.
Since~$R$ and \eqref{eq:r1x.2} are strongly equivalent, it follows that 
SM$_{p_1, \dots, p_k}[F]$ and SM$_{p_1, \dots, p_k}[F']$ are equivalent second-order formulas.
Similarly, we can minimize the scope of the quantifiers in~\eqref{eq:r2x} which will result in the following rule 
\beq
{\forall}{\bf y}\left (  (\exists {\bf x}\; {\alpha({\bf x,y}))}\rar 
p_{k+1}({\bf y})\right).
\eeq{eq:r2x.2}
%where {\bf y} denotes all variables in $K$ that are not in {\bf x} (and recall that
% the variables ${\bf x}$ are being projected.)
Since the transformation from~\eqref{eq:r2x} to~\eqref{eq:r2x.2} is  
intuitionistically equivalent, it follows that $${\text SM}_{p_1, \dots, 
p_{k+1}}[G]$$ is equivalent to
\beq
%\ba{l}
\hbox{SM}_{p_1, \dots, p_{k+1}}[\Gamma\wedge%\\ ~~~~~~~~~~~~
\widetilde{\forall}\left ( \left ( p_{k+1}({\bf y}) \land \beta \right ) 
\rar \gamma \right )\wedge%\\ ~~~~~~~~~~~~
{\forall}{\bf y}\left (  (\exists {\bf x} {\alpha({\bf x,y}))}\rar 
p_{k+1}({\bf y})\right)
]
%\ea
\eeq{eq:grev}
where $\Gamma$ is the conjunction of rules in $F$ other than~$R$. %rules \eqref{eq:r1x} and \eqref{eq:r2x}.
It is sufficient to show that \eqref{eq:grev} is a 
conservative extension of  
SM$_{p_1, \dots, p_k}[F']$.
%{\em First claim to show.} 
Let $M$ be a model of SM$_{p_1, \dots, p_k}[F']$. We will show that  we can construct an 
interpretation $M'$ that coincides with $M$ on the symbols in $\sigma(F')$ and 
is a model of~\eqref{eq:grev}. 
We construct $M'$ %over the signature of~\eqref{eq:grev} 
such that 
\begin{itemize}
\item it coincides with $M$ on all of the symbols in $\sigma(F')$  and 
\item it interprets $p_{k+1}$ so that the following equivalence is satisfied 
\beq
{\forall}{\bf y}\left (  (\exists {\bf x}\; {\alpha({\bf x,y}))}\lrar 
p_{k+1}({\bf y})\right).
\eeq{eq:teq}  
\end{itemize}
%It is easy to see that $F'$ is of the form 
%$$\Gamma\wedge \widetilde{\forall}\left ( \left ( (\exists {\bf x}\;\alpha({\bf x,y})) \land \beta \right ) \rar \gamma \right ),$$
It is easy to check that SM$_{p_1, \dots, p_k}[F']$ is the conjunction of the formulas 
\beq
\Gamma\wedge \widetilde{\forall}\left ( \left ( (\exists {\bf x}\alpha({\bf x,y})) \land \beta \right ) \rar \gamma \right )
\eeq{eq:smfpr1}
and
\beq
%\begin{align}
\ba{ll}
\neg\exists{u_1, \dots, u_k}(&(u_1, \dots, u_k<p_1, \dots, p_k)\wedge \\
&\Gamma^*(u_1, \dots, u_k) \wedge \\
& \widetilde{\forall}\left ( \left ( 
(\exists {\bf x}\;\alpha({\bf x,y})) \land \beta \right ) \rar \gamma \right)\wedge\\
&\widetilde{\forall}\left ( \left ( 
(\exists {\bf x}\;\alpha({\bf x,y})^*(u_1, \dots, u_k)) \land \beta^*(u_1, 
\dots, u_k) \right ) \rar \gamma^*(u_1, \dots, u_k) \right)).
\ea %\end{align}
\eeq{eq:smfpr2}
Formula~\eqref{eq:grev} is the conjunction of the formulas 
\beq
\Gamma\wedge 
\widetilde{\forall}\left ( \left ( p_{k+1}({\bf y}) \land \beta \right ) \rar \gamma 
\right )\wedge{\forall}{\bf y}\left (  (\exists {\bf x}\; {\alpha({\bf x,y}))}\rar 
p_{k+1}({\bf y})\right)
\eeq{eq:smgpr1}
and
\begin{align}
 \neg\exists{u_1, \dots, u_{k+1}}(& (u_1, \dots,u_{k+1}<p_1, \dots, p_{k+1})\wedge\label{exp1}\\
&\Gamma^*(u_1, \dots, u_{k}) \wedge\label{exp2}\\ 
&\widetilde{\forall}\left ( \left ( p_{k+1}({\bf y}) \land \beta \right ) \rar \gamma \right)\wedge\label{exp3}\\
& \widetilde{\forall}\left ( \left ( u_{k+1}({\bf y}) \land \beta^*(u_1,\dots,u_k) \right ) \rar \gamma^*(u_1,\dots,u_k) \right)\wedge\label{exp4}\\
&{\forall}{\bf y}\left (  (\exists {\bf x}\; {\alpha({\bf x,y}))}\rar 
p_{k+1}({\bf y})\right) \wedge\label{exp5}\\
&{\forall}{\bf y}\left (  (\exists {\bf x}\; {\alpha({\bf x,y})^*(u_1,\dots,u_k))}\rar 
u_{k+1}({\bf y})\right))\label{exp6}.
\end{align}

Note that since $\Gamma$ has no occurrences of $p_{k+1}$, $\Gamma^*(u_1, \dots, 
u_{k+1})$ and $\Gamma^*(u_1, \dots, u_{k})$ are identical, and similarly for 
$\alpha^*$, $\beta^*$, and  $\gamma^*$. Expressions~(\ref{exp2},\ref{exp4},\ref{exp6}) reflect this observation.

We now introduce some additional notation required to state the proof.
Let $\mathcal{U}$  denote the universe of interpretation $M'$  (which is also the universe of $M$).
For predicate symbol $q$ and interpretation $I$, let $q^{I}$ denote the function assigned to $q$ by~$I$.
For a formula $H$, let $H^I$ denote the truth value assigned to $H$ by interpretation~$I$.  

% For a tuple element $w$ in $\bu,v$ by $w_\dagger$ we will denote the element in 
%$\bp,t$ in the same position (e.g., $v_\dagger$ is $t$). 

It is clear from the construction of $M'$ that if $M$ is a model of 
\eqref{eq:smfpr1} then 
$M'$ is a model of \eqref{eq:smgpr1}.
It remains to show that if $M$ is a model of SM$_{p_1, \dots, p_k}[F']$
then $M'$ is a model of formula~(\ref{exp1}-\ref{exp6}). 

Proof by contradiction. 
Assume $M'$ is not a model of formula~(\ref{exp1}-\ref{exp6}). 
Then % it follows that for tuple $\bu,v=u_1,\dots,u_k,v$ 
there exists a tuple of functions that we denote by
$u^{M'}_1,\dots,u^{M'}_{k+1}$, from $\mathcal{U}^{n(i)}$ (where  $n(i)$ is the arity of predicate variable
$u_i$) to~$\{\false,\true\}$, such that 
\begin{enumerate}
\item\label{ui:cond1} for every $0<i\leq k+1$  the set of tuples 
mapped to $\true$ by the function 
$u_i^{M'}$ is a  subset of the set of tuples mapped to $\true$ by the 
function $p_i^{M'}$, and 
\item\label{ui:cond2} there is some $0<i\leq k+1$ for which the set of tuples 
mapped to $\true$ by the function 
$u_i^{M'}$ is a proper subset of the set of tuples mapped to $\true$ by the 
function $p_i^{M'}$ and 
furthermore, 
\item $M'$ satisfies conjunctive terms \eqref{exp2}--\eqref{exp6}. 
\end{enumerate}

Case 1.
Consider the case when $u_{i}$ $(i< k+1)$ is the element in tuple $u_1, \dots, u_{k+1}$  
for which condition~\ref{ui:cond2} holds.  
We will illustrate that given the set of functions 
$u^{M'}_1,\dots,u^{M'}_{k}$ and interpretation~$M$ all four conjunctive terms of~\eqref{eq:smfpr2} are satisfied. This observation contradicts the assumption that~$M$ is a model of~\eqref{eq:smfpr2} as we found  the set of functions to interpret the predicate variables  $u_1,\dots,u_{k}$ so that all conjunctive terms of~\eqref{eq:smfpr2} are satisfied. 

Conjunctive term 1: By condition~\ref{ui:cond1} and the assumption of this case, 
the functions 
$u^{M'}_1,\dots,u^{M'}_{k}$ are such that the conjunctive term $(u_1, \dots, u_k<p_1, \dots, p_k)$ of
\eqref{eq:smfpr2} is satisfied by interpretation~$M$.

Conjunctive term 2: Since $M'$ satisfies~\eqref{exp2} when functions 
$u^{M'}_1,\dots,u^{M'}_{k}$ are used to interpret $u_1,\dots,u_{k}$ it follows 
that $M$ satisfies~\eqref{exp2} when the same functions are used to interpret 
$u_1,\dots,u_{k}$. 
(Note,~$\Gamma$ has no occurrence of $p_{k+1}$). Expression~\eqref{exp2} is the second conjunctive term of~\eqref{eq:smfpr2}. 

Conjunctive term 3: Since  $p_{k+1}^{M'}=(\exists 
{\bf x}\;\alpha({\bf x,y}))^{M'}=(\exists 
{\bf x}\;\alpha({\bf x,y}))^{M}$ (following from the construction of~$M'$ and 
the fact that $\exists 
{\bf x}\;\alpha({\bf x,y})$ is over signature of $M$) and since~$M'$ satisfies~\eqref{exp3}, 
it follows that~$M$ satisfies the third conjunctive term of~\eqref{eq:smfpr2}.

Conjunctive term 4: From the fact that $M'$ 
satisfies~\eqref{exp4} and~\eqref{exp6} when functions 
$$u^{M'}_1,\dots,u^{M'}_{k+1}$$ are used to interpret $u_1,\dots,u_{k+1}$ and the 
fact that the fourth conjunctive term of~\eqref{eq:smfpr2} has no occurrence of $u_{k+1}$ or $p_{k+1}$, it follows that 
$M$ satisfies the fourth   conjunctive term of~\eqref{eq:smfpr2} when the same 
functions are used to interpret $u_1,\dots,u_{k}$.

Case 2.
Consider the case when $u_{k+1}$ is the element in tuple $u_1, \dots, u_{k+1}$  
for which condition~\ref{ui:cond2} above holds.  Consider 
a tuple $\Theta$ in $\mathcal{U}^n$ (where $n$ is arity of $p_{k+1}$) so that
$p_{k+1}^{M'}$ maps $\Theta$ to $\true$, while $u_{k+1}^{M'}$ maps~$\Theta$ 
to $\false$. By the construction of $M'$ we know that $p_{k+1}^{M'}=(\exists 
{\bf x}\;\alpha({\bf x,y}))^{M'}$. 
%By~\ref{cexp5}, 
From the last two sentences and the fact that  $M'$ satisfies \eqref{exp5} it follows that 
\beq
(\exists {\bf x}\; \alpha({\bf x},\Theta))^{M'}=\true
\eeq{eq:condcontr}
%For condition~\ref{cexp6} to hold 
To satisfy \eqref{exp6}
for the case of tuple $\Theta$ given that $u_{k+1}^{M'}(\Theta)=\false$, the 
condition $$(\exists {\bf x} \; \alpha({\bf x},\Theta)^*(u_1, \dots, 
u_k))^{M'}=\false$$ must hold. 

Case 2.1. The expression $\alpha({\bf x},\Theta)$ contains no predicate symbols $p_1,\dots,p_{k}$.
Then $\alpha({\bf x},\Theta)$ and $\alpha({\bf x},\Theta)^*(u_1, \dots, u_k)$ coincide.
Recall that condition~\eqref{eq:condcontr}  holds. It follows that this case is impossible.

Case 2.2. The expression $\alpha({\bf x},\Theta)$ contains symbols from $p_1,\dots,p_{k+1}$.

Case 2.2.1. For every symbol $p_i$ in $\alpha({\bf x},\Theta)$, it holds that $p_i^{M'}= u_i^{M'}$.
It follows that  $$(\exists {\bf x}\; \alpha({\bf x},\Theta))^{M'}=(\exists {\bf x}\; \alpha({\bf x},\Theta)^*(u_1, \dots, u_k))^{M'}.$$ Recall that condition~\eqref{eq:condcontr}  holds. It follows that this case is impossible.

Case 2.2.2. For some symbol $p_i$ in $\alpha({\bf x},\Theta)$, it holds that 
the set of tuples 
mapped to $\true$ by the function~$u_i^{M'}$ is a proper subset of the set of tuples mapped to $\true$ by the 
function $p_i^{M'}$. Note that $i<k+1$ as $p_{k+1}$ does not occur in $\alpha(\cdot,\cdot)$. The argument of Case 1 applies.

{\em Second claim to show.} (Illustration of 1-1-correspondence) We have to show that given an interpretation~$M$ of SM$_{\bf p}[F']$, $M'$ constructed in  the first claim  is the only interpretation that is a model of~\eqref{eq:grev}  
and that coincides on symbols in $M$. This claim follows from Theorem 10 from \cite{fer09}.

{\em Third claim to show.} Given a model of~\eqref{eq:grev} show that it is a model of SM$_{\bf p}[F']$. This is a simple direction. E.g., by contradiction.
\end{proof}
%\end{comment}
\end{document}